\documentclass[]{article} 
\usepackage{emulateapj}
\usepackage{graphicx}
\usepackage{multirow}

\advance \voffset by -0.5cm\relax

\def\msun{{\rm ~M}_{\odot}}

\def\mps{{\rm ~M}_{\odot} {\rm ~s}^{-1}}

\begin{document}

\title{Missing Black Holes Unveil The Supernova Explosion Mechanism}

\author{Krzysztof Belczynski\altaffilmark{1,2}, Grzegorz Wiktorowicz\altaffilmark{1}, 
        Chris L. Fryer\altaffilmark{3}, Daniel E. Holz\altaffilmark{4,5},
        Vassiliki Kalogera\altaffilmark{6}}

 \affil{
     $^{1}$ Astronomical Observatory, Warsaw University, Al.
            Ujazdowskie 4, 00-478 Warsaw, Poland (kbelczyn@astrouw.edu.pl)\\
     $^{2}$ Center for Gravitational Wave Astronomy, University of Texas at
            Brownsville, Brownsville, TX 78520\\
     $^{3}$ Computational Computer Science Division, Los Alamos National
            Laboratory, Los Alamos, NM\\
     $^{4}$ Enrico Fermi Institute, Department of Physics, and Kavli Institute
            for Cosmological Physics, University of Chicago, Chicago, IL 60637\\
     $^{5}$ Theoretical Division, Los Alamos National Laboratory,
            Los Alamos, NM 87545\\
     $^{6}$ Center for Interdisciplinary Exploration and Research in Astrophysics 
            (CIERA) \& Dept. of Physics and Astronomy, Northwestern University, 
            2145 Sheridan Rd, Evanston, IL 60208
}
 
\begin{abstract}
It is firmly established that the stellar mass distribution is smooth,
covering the range $0.1$--$100\msun$. It is to be expected that the masses
of the ensuing compact remnants correlate with the masses of their
progenitor stars, and thus it is generally thought that the remnant masses
should be smoothly distributed from the lightest white dwarfs to the
heaviest black holes. However, this intuitive prediction is not borne out
by observed data. In the rapidly growing population of remnants with
observationally determined masses, a striking mass gap has emerged at the
boundary between neutron stars and black holes. The heaviest neutron stars
reach a maximum of two solar masses, while the lightest black holes are at
least five solar masses. Over a decade after the discovery, the gap has
become a significant challenge to our understanding of compact object
formation. We offer new insights into the physical processes that
bifurcate the formation of remnants into lower mass neutron stars and
heavier black holes. Combining the results of stellar modeling with
hydrodynamic simulations of supernovae, we both explain the existence of
the gap, and also put stringent constraints on the inner workings of the
supernova explosion mechanism. In particular, we show that core-collapse
supernovae are launched within $100$--$200$ milliseconds of the initial
stellar collapse, implying that the explosions are driven by instabilities
with a rapid (10--20\,ms) growth time. Alternatively, if future
observations fill in the gap, this will be an indication that these
instabilities develop over a longer ($>200$ milliseconds) timescale.
\end{abstract}


\section{Introduction}

Our Universe is littered with black holes, ranging from ones roughly the
mass of our Sun to behemoths many million times more massive. Understanding
how these black holes are formed, and how many there are over the course of
the Universe's history, is one of the defining challenges of modern
astrophysics. The formation of black holes is related to the life and death
of stars, involving stellar evolution, accretion, stellar explosions, binary
formation, galaxy feedback, and a host of other important physical
processes. In the past decade advances have occurred both in the observation
of black hole systems and in the theory underlying them. Although the
formation of the super-massive black holes remains poorly understood, there
is a growing consensus on the formation and evolution mechanisms for black
holes born out of dying stars. Our focus is on these stellar mass black
holes. 

A total of about $50$ stellar-mass black hole systems (Ziolkowski 2010) and 
about 1,000 neutron stars have been observed in the local Universe (Liu et
al. 2006). Of these systems, we have mass determinations for about $20$ black 
holes and $50$ neutron stars. The most striking feature of the mass 
distribution of these compact objects is the observed gap in remnant masses 
between neutron stars and black holes. As first noted by Bailyn et al. (1998), 
there are no observed compact remnants in the mass range $2$--$5\msun$. 
The most recent measurements and independent sophisticated statistical analyses 
have further established this ``mass gap'' (Ozel et al. 2010; Farr 
et al 2011). Exploring this mass range is critical to understanding the 
equation-of-state of dense nuclear matter that is otherwise inaccessible, 
as well as black hole formation and the supernova engine. We investigate the 
formation of neutron stars and black holes to show that the mass gap can be 
naturally explained, and that its existence puts strong constraints on the 
underlying supernova engine.

\section{Supernova Explosions}

There is general agreement that the supernova engine is powered by the collapse
of a massive star. To produce an explosion, the star must eject its outer
layers. Although there exist a number of physical processes for initiating the
explosion, there is consensus among supernova modelers that instabilities play 
an important role in increasing the efficiency with which the gravitational 
energy from the collapsing star is tapped to drive an explosion. Stars in the 
mass range $\sim 10$--$20\msun$ experience strong supernova explosions and 
produce low-mass neutron stars. Most massive stars ($\gtrsim 40\msun$) fail to 
explode, and instead form massive black holes. It is generally accepted that 
stars with masses in-between undergo weak supernovae, producing both 
higher-mass neutron stars and lower-mass black holes. This would produce a 
continuous distribution of remnant masses, including some within the mass gap, 
and is therefore in tension with current observations. Although differing 
models of instability introduce different mass distributions within the gap 
region, none adequately reproduce the observed gap. In the current work we show 
that, by positing a sufficiently rapid growth of turbulence in the supernova 
explosion engine, combined with a detailed binary star evolution model, the 
observations can be fully explained.

Neutron stars and black holes are thought to form from core-collapse
supernovae. Initial core collapse is halted when a proto-neutron star 
forms, the infalling material ``bounce'' off of the suddenly rigid 
core, and the outgoing shock runs into the rapidly infalling outer 
layers of the star, and eventually stalls. Some fraction of a strong 
flux of neutrinos from proto-neutron star is absorbed and heats the 
``turbulent atmosphere'', the layer between the proto-neutron star 
and the stalled shock. A supernova engine is successful only if this 
energy can be utilized to revive the shock, allowing it to overcome 
the pressure from the infalling material. For sufficiently strong 
convective motions, the infalling layers may be pushed out, and the 
explosion may be restarted. However, an initial jolt is required to 
start convection.  
In the past two decades it has been realized that the turbulent region
plays a crucial role in accumulating the energy from collapse, and
that an instability must then trigger the convective engine leading to
the supernova explosion (Herant et al 1994; Fryer \& Warren 2002;
Blondin et al. 2003; Burrows et al.  2006; Fryer \& Young 2007; Scheck
et al. 2008; Bruenn et al. 2009; Marek \& Janka 2009; Nordhaus et al. 
2010; Takiwaki et al. 2011).  
At present the physical nature of the force initiating the explosion 
is poorly understood, and the precise mechanism for the instability and 
shock revitalization are under active study. As we discuss below, the 
differing models can be broadly characterized by their growth times.  
Two models that demonstrate the differences in the growth times are the 
 and standing accretion shock instabilities, the former 
having a growth time of nearly a factor of 10 shorter than the latter. 
In this paper we seek to distinguish between these growth times by 
studying two extreme models: a rapid instability model assuming a 
10--20\,ms growth time, and a delayed instability model assuming a 
100--150\,ms growth time. For simplicity, we dub these models as rapid 
and delayed, respectively.

In a collapsing star, the Rayleigh-Taylor is encountered just above the 
surface of the proto-neutron star. Neutrinos heat the turbulent 
atmosphere from below creating a temperature gradient, while the 
infalling gas from above creates a density gradient. As a result, low 
entropy gas finds itself on top of high entropy material, and a violent 
displacement of layers follows. Once the movement of the plasma is 
initiated, it turns into a convective engine. For the conditions of the 
post-shock stall region, Fryer \& Young (2007) found that the 
Rayleigh-Taylor instability may appear very early, at $\sim 20$\,ms after 
the initial bounce. However, the convective engine needs additional time 
($\gtrsim 100$\,ms) to build up sufficient strength to fully move the medium
above and launch the explosion. If this engine is strong enough to drive an 
explosion, the explosion will happen rapidly. In simulations, if this 
instability does not launch the explosion within the first $200$--$300$\,ms, 
it means that the mechanism has failed, and the star will not explode. We 
note, however, that not all supernova models lead to explosions through the
development of the Rayleigh-Taylor instability at early times. The 
differences in model outcomes may be due to differing treatments of neutrino 
transport and/or hydrodynamics (e.g., numerical viscosity).

The standing accretion shock appears where the initial bounce stalls, and 
gives rise to additional instabilities. One example is the vortical-acoustic 
instability that arises and is amplified by the standing shock. Some of the 
outer stellar material is essentially in free fall as the star collapses. 
These blobs of infalling matter can acquire significant momentum by the time 
they hit the standing shock, and the resulting ``pounding'' causes the 
underlying plasma entrapped in the turbulent atmosphere to vibrate and 
create pressure waves. The interference of these waves transfers momentum to 
the plasma, which in turn starts the convective engine. This is a heuristic 
description; the full numerical ``discovery'' and inner workings of these 
instabilities is reviewed by Foglizzo (2009) Note that in the case standing 
accretion shock instabilities the initial perturbation that triggers the 
convective engine comes from above (as opposed to be sourced just above 
proto neutron star surface in the Rayleigh-Taylor instability). Blondin et al. 
(2003) estimate the growth time of this ``acoustic'' instability to be 
considerably longer than the Rayleigh-Taylor case; strong standing accretion 
shock instabilities require $\gtrsim 500$\,ms to develop, and may result in 
supernova explosion as late as $1000$\,ms after the bounce.
 
With time the proto-neutron star cools off, and the deposition of energy into 
the turbulent region declines. The turbulent atmosphere is also subject to 
cooling, and eventually the energy stored in this region begins to decline. 
The amount of energy that can be extracted from this region is therefore 
dependent on how long it takes before the convective engine turns on. 
The Rayleigh-Taylor is the first instability to appear. If it is strong enough 
to start the convection {\em  and}\/ if there is enough energy accumulated in 
the turbulent region by that time to eject the outer layers of the star, a 
rapid explosion follows (within $\lesssim 200$--$300$\,ms). If these conditions 
are not met, then the infall continues, leading to the development of standing 
accretion shock instabilities. If these instabilities are strong enough a 
delayed explosion follows (after $\sim 500$--$1000$\,ms). If neither scenario
comes to pass, the star fails to explode, and infall continues until the 
proto-neutron star collapses into a black hole. 

Supernova modelers debate the precise emergence times and energy outputs of 
the different instabilities, in an effort to identify which are strong enough 
to trigger the convective engine. The answer depends on detailed numerical 
treatment of the turbulent region, as well as on highly uncertain physics of 
extremely dense and hot plasma (convection, neutrino transport, radiation-matter 
interactions). Obviously, this is the key issue in a quest to understand 
supernova explosions. As the appropriate underlying physics, and ensuing 
instabilities, continue to be contested, we have decided to subsume our 
ignorance of the detailed physical mechanism, and instead parameterize the 
appropriate physics through the turbulent growth time. We find that this growth 
time directly relates to the masses of compact remnants. Comparing the predicted 
mass distributions for both short and long growth times with the observed masses 
of Galactic neutron stars and black holes, we find that only an instability that 
develops within the first $\sim 10-20$\,ms after the bounce and leads to a 
rapid explosion ($\sim 100-200$\,ms) can account for the data. Any mechanism 
that can grow and drive the explosion on this timescale can be consistent with 
the observed mass gap. 
 
The nature of the infalling stellar material depends upon the density structure 
of the star prior collapse. Just before collapse the center, which determines 
the fate of the star, is composed of iron covered by silicon and oxygen layers. 
The extent of these layers depends sensitively on the mode of energy transport 
while these layers were being formed. For relatively low temperatures (lower mass 
stars) the energy output from nuclear fusion is modest, and the energy is transported 
radiatively. or high temperatures (high mass stars) the energy output is
sufficiently high that convection turns on. The convective mixing drags additional 
fuel into the burning zones, extending the lifetime of the star and producing more 
massive silicon and oxygen layers of high density. Although the exact limit depends 
sensitively on details of nuclear burning and is poorly established, the transition 
from radiative to convective burning is very abrupt in terms of star mass of about 
$20$--$25\msun$ (Woosley et al. 2002). 

The inner part of the stellar core collapses to an $\sim 1 \msun$ proto-neutron star. 
After the bounce this dense object accretes from the turbulent region at extremely high 
rates (as high as $\sim 1 \mps$). Depending on the delay before the explosion, the mass 
of the proto-neutron star may increase via post-bounce accretion by up to $\sim 1
\msun$. After the explosion accretion via fallback is encountered. Even for strong 
explosions some amount of fallback is expected ($\sim 0.1$--$0.2 \msun$). For weaker 
explosions the ejected material has less kinetic energy, and more fallback is noted.  

In the initial mass regime for core collapse supernovae ($M \sim 8$--$14 \msun$), the
density of the star falls off steeply outside the very center. In addition, the early 
energy deposition is very efficient, and the energy accumulated in the turbulent region 
increases rapidly. For these stars the convective engine isn't necessary, as the strong 
flux of neutrinos heats the material above the proto-neutron star, and the resulting 
pressure is able to drive the explosion. The convective engine is initiated early by 
the Rayleigh-Taylor instability, and may enhance the explosion energy, resulting in 
strong to moderately strong explosions. As there is little time for post-bounce 
accretion and additionally very little fallback is expected, low mass neutron stars 
are formed: $\sim 1$--$1.5 \msun$. For higher mass stars the compact remnant formation 
process is different for the delayed and rapid models, as we detail below. 

For stars in the low mass regime ($M \sim 14$--$20 \msun$) the supernova explosions 
are very strong, owing to the moderate density of infalling material. The infalling 
material can hold off the explosion for long enough to allow significant energy build 
up in the turbulent region, and yet this pressure lid is not strong enough to delay 
the explosion to the point that the energy in the turbulent region starts to decline.
Because the pressure of the infalling material decreases with time, the explosions in 
the rapid model are more energetic than in the delayed model. As a consequence, high 
mass neutron stars are formed in the rapid model: $\sim 1.5$--$2 \msun$ with 
significant post-bounce accretion and rather weak fallback. In the delayed model, fall 
back becomes more pronounced and compact objects with mass extending to higher values 
$\sim 1.5$--$3 \msun$ are predicted. 

In the intermediate mass regime ($M \sim 20$--$40 \msun$) extended layers of high 
density are present. In the rapid model, the infalling silicon/oxygen layers prevent 
the convective engine to lunch the explosion. There is no (or almost no) explosion, 
and the entire star collapses onto the proto-neutron star, followed by the formation of 
a black hole. Depending on the mass of the exploding star (mostly a function of initial 
stellar mass and the strength of stellar winds) black holes with mass $5$--$10 \msun$ 
are formed. In the case of the delayed model, however, by the time the standing shock 
instabilities arise to drive the convective engine, the high density layers are 
finishing their infall. Since the infalling layers are now lower density, the
pressure lid is weak, and any resulting explosion is similarly weak. Depending on the 
extent of the silicon/oxygen layers, compact objects within the mass range $3$--$5 \msun$ 
are formed. The range results from more massive pre-supernova stars leading to weaker 
explosions and increased fallback.

For the most massive stars ($M > 40 \msun$) the outcome of core collapse is 
independent of the supernova model. These stars have very extended high density 
layers in their centers. In the case of either the early or delayed instabilities, the 
convective engine releases insufficient energy to overcome the high pressure of 
the infalling material. Even if a weak explosion manages to occur, all of the  ejected 
material is subject to fallback, and the entire star is accreted onto a compact object 
followed by the formation of a massive black hole ($5$--$15 \msun$). 

For the delayed supernova model, the explosions range from weak to strong,
resulting in a wide spectrum of compact object masses. In particular, no gap in
mass is predicted. In the rapid supernova model, on the other hand, the
explosions are either strong, or fail. The break in stellar evolution, from radiative 
to convective burning, coupled with the rapid rise of the supernova engine, leads to a 
commensurate break in the masses of the compact remnants. In practice, the mass
gap in the rapid model is not as dramatic as presented in our simplified arguments 
above. For example, some of the rapid explosion models predict a small number of 
compact objects within the range $2$--$5 \msun$ (e.g. Fryer et al. 2012). These 
compact objects would be born out of the weakest of the rapid explosions, launched 
just as the Rayleigh-Taylor instability fizzles out (at $200$--$300$\,ms after the 
bounce). Guided by the observed mass gap, in our models we have limited the time 
of explosion to $100$--$200$\,ms, which precludes any compact object remnants in
the $2$--$5\msun$ mass range. This condition was imposed in the rapid model version 
of our description of single stellar evolution, and then further employed in our 
binary star analysis. The (very few) objects within the mass gap found in the rapid 
model are the result of accretion onto compact objects from their binary
companions. Since all known black holes are found in binary systems, we now  
turn to binary evolution to test whether the mass gap predicted for single stars
prevails in binary populations.

\section{Binary Evolution}

We have employed a Monte Carlo ``population synthesis'' method to
follow the evolution of several million binary stars from their birth in 
the gravitational collapse of gas clouds, through 10 billion years of Galactic 
history until the present. Specifically, we have implemented two major 
supernova explosion scenarios in the {\tt StarTrack} population synthesis 
code developed by Belczynski et al. (2002). The two scenarios 
provide strikingly different mass distribution for massive remnants in the case 
of single star evolution. In particular, Fryer et al. (2012) find that 
the rapid explosion model is depleted of remnants in the mass range
2--5$\msun$, while the delayed model, utilizing explosions sourced by the
standing accretion shock instability, delivers a continuous mass distribution
from neutron 
stars through to black holes, with no gap in mass. However, since all the known black
holes are 
found in X-ray binary systems, we ask whether the results of Fryer
et al. (2012) are sustained in binary populations. Various binary interactions
involving  mass transfer episodes between binary component stars may increase
the mass of neutron stars in the rapid model, and thereby may wash out the gap. 
Alternatively, other evolutionary processes may severely deplete the number
of binaries that host low mass black holes, and produce a gap in the
delayed model. An example would be a selective increased disruption of binaries 
with low mass black holes via natal kicks caused by supernova asymmetries 
(e.g., Hobbs et al. 2005). 

Our code incorporates the major physical processes to be expected
in the evolution of binaries with neutron stars and black holes. 
In particular, various modes of mass transfer/loss are followed in
detail (Belczynski et al. 2008a), the spin evolution and accretion onto
compact objects is accounted for (Belczynski et al. 2008b) and recent estimates
for stellar mass loss rates (Belczynski 
et al. 2010) and natal kicks were also included (Fryer et al. 2012). 
The {\tt StarTrack} remnant mass distribution was based on 
standard stellar models (Hurley, Pols \& Tout 2000) interposed with 
previous generation supernova simulations (Timmes, Woosley \& Weaver 1996) ,
and accounts for a range of explosion energies and fallback estimates 
(Fryer \& Kalogera 2001).

We find that the major features of both models in the single star case are
preserved in binary 
populations. Figure~1 shows the current Galactic population of 
X-ray binaries found in our evolutionary Monte Carlo simulations. As expected,
both the rapid and delayed models are dominated by neutron star 
X-ray binaries (remnants with mass smaller than $2\msun$), as stars forming 
these remnants are more abundant than the more massive progenitors of black 
holes (e.g., Kroupa \& Weidner 2003). 
For massive remnants the models make distinct predictions.
In the rapid supernova scenario, we find 
few to no Galactic X-ray binaries with remnants in the mass range 
2--5$\msun$. By contrast, the delayed model allows for a significant population
of X-ray binaries in this mass range. Both models provide
X-ray binaries with black holes more massive than 5$\msun$, extending to about
$15\msun$ and coinciding with the most 
massive black holes observed in the Galaxy: Cyg X-1 ($14.8\msun \pm 1.0$,  
Orosz et al. 2011) and GRS 1915 ($14\msun \pm 4$, Casares 2007).

A continuous compact object mass distribution is to be expected from the delayed
model, since a decreasing supernova energy results in larger fallback and larger 
compact object mass with increasing progenitor mass. The 
emergence of a gap in the rapid model occurs at compact object mass of $\sim 2 
\msun$ for the evolution of a single star (Fryer et al. 2012). A compact object
of such a mass is formed out of a $M_{\rm zams} \sim 20$--$25 \msun$ progenitor 
(see, e.g., Belczynski et al. 2008a). 
As seen from Figure~1, the lower-mass gap boundary starts at $\sim 3 \msun$
if the effects of binary evolution are included. It is 
still a subject of debate whether compact objects with mass in the 
range 2--3$\,\msun$ are NSs or BHs. The only known system with a dynamical mass 
estimate in this range is 4U 1700-37 ($M=2.44 \pm 0.27$: Clark et al. 2002; 
$M=2.58 \pm 0.23$: Rude et al. 2010, J.Orosz 2011, private communication). The 
nature of the compact object in 4U 1700-37 is not yet established, although 
recent detection of quasi-periodic oscillations (Dolan 2011) indicates a low mass 
BH as originally suggested by Brown, Weingartner, \& Wijers (1996). For initial 
progenitor masses higher than $M_{\rm zams} \sim 20$--$25 \msun$, the supernova
models based on  
Rayleigh-Taylor instability fail to explode (e.g., Fryer et al. 2012). More 
massive progenitors collapse to form heavy BHs. For example, a $M_{\rm zams} = 30 
\msun$ single progenitor collapses to a $\sim 10 \msun$ BH (the rest of the progenitor mass 
is lost during its nuclear evolution via stellar winds). At $M_{\rm zams} \gtrsim 
35 \msun$ single progenitors are subject to enhanced mass loss via Luminous
Blue  
Variable (LBV) winds (e.g., Vink \& de Koter 2002). As a result, the high mass 
progenitors at core collapse may be less massive than the initially lighter 
stars. For example, a $M_{\rm zams} \sim 40 \msun$ progenitor collapses to a $5-6 
\msun$ BH. For the most massive single progenitors $M_{\rm zams} \gtrsim 80 \msun$ the 
combined effect of the very high progenitor mass and the shortness of the LBV 
phase (stars at this mass rapidly evolve into a W-R stage that is not a subject to 
the LBV-type mass loss) allows for the formation of BHs with mass 10--15$\,\msun$. 
The onset of the LBV phase marks the formation of the lightest BHs in the rapid 
model and terminates the mass gap at the compact object mass of $5-6 \msun$
in the case of single stellar evolution. 
In binary evolution, the lowest mass BHs are formed either through onset of
the LBV phase (non-interacting binaries) or via combination of the LBV phase and 
the common envelope evolution (close binaries; e.g., Webbink 1984). Both processes, 
although very different in their nature, generate the same outcome: H-rich envelope 
removal that leads to the formation of the lowest mass black holes in the rapid 
model at $5-6 \msun$.

The gap, or lack thereof, is the primary signature that distinguishes 
the two supernova models, and it originates directly from the physics
outlined earlier in the text and based on the evolutionary and supernova models 
compiled by Fryer et al. (2012).
From the binary evolution perspective, mass accretion onto neutron stars is 
insufficient to wash out the gap in the rapid model, nor does it produce a
gap in the delayed model (e.g., from natal kicks) where none was present before. 

In the case of low mass X-ray 
binaries, although prolonged mass transfer episodes may be expected, a low 
mass companion ($\lesssim 1 \msun$) does not provide enough of a mass reservoir 
to increase the neutron star mass significantly over $2 \msun$.
For high mass X-ray binaries, there are enormous amounts of mass in the
massive companion stars. However, since in these cases the mass ratio of both 
components is extreme (typically a $1.4 \msun$ neutron star and a $\gtrsim
5$--$10 \msun$ 
bright companion star) the mass transfer via Roche lobe 
overflow is sufficiently fast and violent that most of the mass is ejected from
the binary  
instead of being accreted onto the neutron star (e.g., Tauris \& Savonije 
1999; Dewi \& Pols 2003). This follows from the 
Eddington limit, where photon 
emission generates a strong countervailing wind and inhibits further
accretion. In both of these cases the accretion from the companion's stellar
wind does not become a significant factor in increasing a neutron star's mass,
although it may drive a significant X-ray luminosity, as observed in some 
wind-fed X-ray binaries (e.g., Liu, van Paradijs \& van den Heuvel 2006).

On the other hand, the formation channels providing X-ray binaries do not 
selectively prevent the formation of systems with low mass black holes.
Natal kicks are believed to be highest for neutron stars; the progenitors
are almost fully disrupted in supernova explosions, and thus are characterized by 
significant asymmetries. With increasing mass of the progenitor, the supernova 
energetics drops and the matter initially ejected falls back onto the proto 
compact object, forming a black hole. In this case moderate asymmetry is expected. Finally, 
for the most massive stars the entire star collapses to form a black hole, and 
little asymmetry (and thus kicks) is expected (e.g., Fryer \& Kalogera 2001; 
Mirabel \& Rodrigues 2003). For remnant masses in the mass range 2--5$\msun$ 
(most likely light black holes if they exist), the predicted fallback is 
significant. We find that the resulting kicks are therefore insufficient to 
deplete the X-ray binary population and they cannot create the mass gap in the 
delayed model. 

We note that Figure~1 only includes Galactic stellar population models assuming
solar metallicity ($Z_\odot=0.02$). Due to rather high wind mass loss rates from
stars at this metallicity (e.g., Vink, de Koter \& Lamers 2001) the maximum BH
mass, both observed and predicted in our models, is only $\sim 15
\msun$. However, if the metallicity is decreased the mass loss rates drop as
well, and the maximum BH mass can extend to $\sim 80 \msun$ in our models for
$Z=0.01 Z_\odot$ (Belczynski et al. 2010). In particular, the most massive known
black hole of stellar origin BH IC10 X-1 with mass of $\sim 30
\msun$ (Prestwich et al. 2007) is naturally explained by our models (Bulik, Belczynski, \& 
Prestwich 2011). The high end of the compact object mass spectrum is 
insensitive to the supernova explosion engine, as the most massive stars go
through very weak explosions and the entire (or almost entire) immediate
progenitors end up as compact object. This is why the rapid and delayed models
look virtually identical for compact-object masses higher than $6 \msun$ (see
Fig.~1).  The major limiting factor on the maximum BH mass, for a given
metallicity, is the efficiency of stellar winds in removing mass from the
progenitor star. The winds, in turn, are coupled to the driving radiative force
and thus are set by stellar evolution (luminosity and temperature of BH
progenitor stars). Since empirical estimates of stellar wind rates and the
knowledge of stellar evolution are incomplete, the maximum mass of stellar BH
remains at present an open issue. This, along with uncertainties in accretion
physics, has led to a controversy over whether the ultraluminous X-ray sources are
powered by stellar origin or instead are dynamically formed (intermediate-mass)
BHs (e.g., Farrell 2009; Sutton et al. 2012).

\section{Discussion}

The existence of the mass gap, with an absence of neutron stars and black holes in the 
mass range 2--5$\,M_\odot$, in the current observations is fairly well established. 
At the moment, it cannot be excluded that the gap arises from some potential observational 
biases that may hide low-mass black holes from the current observed sample.  
An example of such a bias is that, in the case of low-mass X-ray binaries powered by 
Roche-love overflow, BH mass measurements are enabled only when the host X-ray binary 
exhibits transient behavior; transient behavior is thought to be suppressed when the mass 
ratio is closer to unity, and therefore it is conceivable that low-mass BHs are biased 
against in the current sample (Fryer \& Kalogera 2001; Ozel et al. 2010). However, in the 
case of wind-fed high-mass X-ray binaries, low-mass black holes would not suffer the same 
bias. 

In a recent study Ugliano et al. (2012) have calculated remnant birth masses
for neutrino driven supernovae explosions. They have performed $1$-D spherically
symmetric hydrodynamical simulations with analytical approximation on a central
proto neutron star and artificially driven explosions. Their study was focused
on solar metallicity stars within initial mass range $M_{\rm zams}=10-40 \msun$.
They have found that the explosions can develop in a broad range of time 
($0.1-1.1$s post bounce) and they have estimated compact object mass spectrum.
They found that {\em (i)} neutron stars form with broad maximum $1.4-1.7\msun$
baryonic mass (that corresponds to $1.3-1.6\msun$ gravitational mass), and
that {\em (ii)} black holes form above $6\msun$ with vast majority of them
found in $12-15\msun$ (baryonic) mass range (see their Fig.6).
Apparently, their findings stand in stark contrast with observational data
on Galactic compact objects (solar metallicity). First, majority of Galactic 
black holes are found with mass $5-12 \msun$ (with just 2 outliers at about 
$\sim 15\msun$: Cyg X-1 and GRS 1915; see our Fig.1) while Ugliano et al. 
(2012) found almost all black holes in range $12-15\msun$. Second, the 
predicted by Ugliano et al. (2012) mass gap starts over $6\msun$ which again 
is in clear contradiction with mass measurement of some black holes with 
masses below that value (e.g., $\sim 5.1\msun$ for a black hole in XTE 
J1650-500; Slany \& Stuchlik 2008). Third, neutron star mass distribution is  
sharply peaked at $\sim 1.35\msun$ (e.g., Lorimer \& McLaughlin 2009), 
although Ugliano et al. (2012) found NS mass evenly distributed over the range 
$1.3-1.6\msun$. Overall, it appears that results obtained by Uglaino et al. 
(2012) are based on too simple of a model (e.g., $1$-D simulations, spherical 
symmetry, not-self consistent but artificially-driven explosions) to provide 
meaningful comparison with observations. Ugliano et al. (2012) major result 
of wide range of explosion times noted to be in contrast with our conclusion 
(only short timescale explosions allowed by current observations) is further 
depreciated by recent $3$-D physically self-consistent supernova simulations 
(e.g., Nordhaus et al. 2010; Murphy, Dolence \& Burrows 2012).

An early premise was that BHs tend to harbor massive companions (e.g., the 
massive O star in Cyg X-1; recently estimated at $19.2 \msun$ by Orosz 2011). 
Brown et al. (1996) estimated that the formation 
rate of X-ray binaries with low- and high-mass BHs are similar. It was
claimed that for low-mass BHs the ensuing Roche-lobe overflow would be very short
(thermal timescale due to extreme mass ratio) and therefore it poses an
observational bias against detection. On the other hand, for binaries with
massive BHs, the accretion proceeds on much longer (e.g., nuclear) timescale and
hence the X-ray phase lasts longer. However, at present the Galactic sample of
known BH binaries consists mostly of transient X-ray sources with low-mass
donors (e.g., Ziolkowski 2010), and hence the premise that the majority of BH
binaries harbors massive companions is no longer supported.

In recent work, Kreidberg et al. (2012) have analyzed in detail the elements
entering the BH mass determination and have identified an important source of a
systematic error that can potentially lead to the masses being overestimated.
Ellipsoidal light curve variations are typically used to establish the 
inclination of a given binary. Most known binaries harboring BHs are interacting 
systems with Roche lobe overflow, and therefore emission from an accretion disk 
and a hot spot are present in the measured flux, even during what is assumed to 
be quiescent phases. These extra contributions of light are often either not 
accounted for or misinterpreted in BH binary light curve analyses. According to 
Kreidberg et al. (2012), this leads to a systematic bias in the inclination 
measurements and hence the BH masses, and may mistakenly push low-mass BHs out 
of the mass gap. Their analysis is based on the high-quality data for A0620-00, 
which they use and extend the analysis for the full sample of BH X-ray binaries 
with low-mass donors. For two of the systems in the sample, GRO J0422+32 and 
4U1543-47, the corrected BH masses turn out to lie within the mass gap. They 
repeat the statistical analysis of Farr et al. (2011) for the corrected
values and find that the statistical significance of a mass gap is eliminated, 
if the corrected BH mass for GRO 0422+32 is indeed correct (the BH mass for
4U1543-47 is burdened with large errors due to limited observations and hence 
does not affect the result in a statistically significant way). Kreidberg et al. 
(2012) note that higher quality observations are needed for both of these 
systems to conclusively assess the existence of a BH mass gap in the current 
sample. 

Nevertheless, we have shown that the observed gap in compact remnant mass
could arise naturally, depending solely on the growth timescale of the
instabilities driving the explosion of massive stars. In fact the observed gap
places strong  
constraints on the development of stellar collapse, with a rapid explosion model
being strongly preferred. This model predicts two distinct fates for a massive
star: either a violent outburst which ejects most of the star and leads to
neutron star formation, or a failed supernova wherein the entire star collapses
to a black hole. For a gap to be present, very few stars can lie in the
intermediate regime where weak explosions occur. An explosive mechanism that
only succeeds within the first $\sim 100$--$200$\,ms produces a remnant mass gap
in accordance with observations. For
slower-growing turbulent instabilities we are unable to match the observed
gap. For the delayed standing accretion shock instabilities to reproduce the
observed mass gap, an extra source of energy is required in the supernova models
(e.g., a magnetar phase) to inhibit the fallback that produces low mass black
holes in weak explosions. Alternatively, if in the future
the mass gap is found to be an observational artifact, and compact remnants are 
found to populate the gap, this will indicate that long growth time, delayed
instabilities occur in supernova explosions. Thus the presence or absence of a
mass gap is a critical clue in unveiling the engine behind supernova explosions.

\acknowledgements
We would like to thank the anonymous referee and Christian Ott for many helpful 
comments. The authors acknowledge support from MSHE grant N203 404939 (GW, KB), NASA 
Grant NNX09AV06A to the UTB (KB), NSF grant AST-0908930 (VK), and from the 
Aspen Center for Physics.

\vspace*{-0.7cm}
\begin{figure}
\includegraphics[width=0.9\columnwidth]{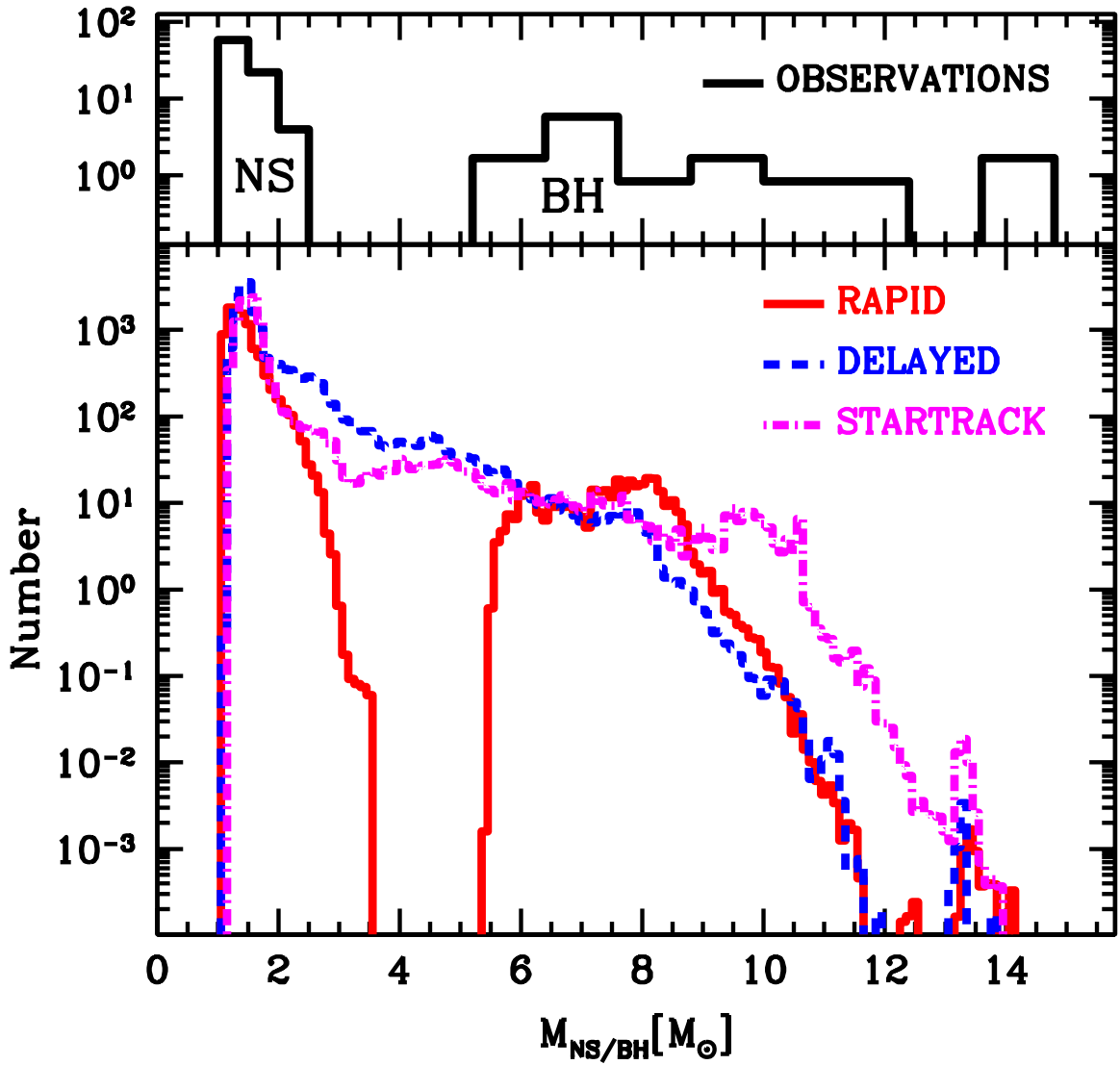}
\vspace*{-1.5cm}
\caption{
Predicted neutron star/black hole mass distribution for the Galactic population 
of Roche lobe overflow and wind fed X-ray binaries. We have employed a population 
synthesis method to generate the binaries with two supernova explosion models. In 
the model in which the explosion is driven on a rapid timescale (RAPID: 
$\lesssim 100-200$\,ms), we note the same striking gap in remnant mass (almost no 
compact objects with mass $2$--$5 \msun$) as found in the observations. 
In contrast, the model in which the supernova explosion is significantly
delayed (DELAYED: $\gtrsim 500-1000$\,ms) shows a continuous compact object mass 
distribution. For comparison, we also show the remnant mass distribution commonly used
in modern studies of Galactic and extra-galactic binaries with compact
objects as implemented in the {\tt StarTrack} population synthesis code. 
It is clear that a major revision in the remnant mass distribution is required
in future studies (e.g., Dominik et al.\ 2012). 
}
\label{all}
\end{figure}


\begin{references}

\reference{} Bailyn, C., Jain, R., Coppi, P., \& Orosz, J.\ 1998, ApJ, 499, 367
\reference{} Belczynski, K., Kalogera, V., Bulik, T.\ 2002, ApJ, 572, 407
\reference{} Belczynski, K., et al.\ 2008a, ApJ Sup., 174, 223
\reference{} Belczynski, K., et al.\ 2008b, ApJ, 682, 474
\reference{} Belczynski K., et al.\ 2010, ApJ, 714, 1217
\reference{} Bruenn, S., et al.\ 2009, AIP Conf., 180, 1 (astro-ph/1002.4914)
\reference{} Burrows, A., et al.\ 2006, ApJ, 640, 878
\reference{} Blondin, J., Mezzacappa, A., DeMarino, C.\ 2003, ApJ, 584, 971
\reference{} Brown, G., Weingartner, J., \& Wijers, R.\ 1996, ApJ, 463, 297
\reference{} Bulik, T., Belczynski, K., \& Prestwich, A.\ 2011, ApJ, 730, 140
\reference{} Casares, J.\ 2007, Proceedings of the IAU Symposium 238, 3
\reference{} Clark, J., et al.\ 2002, A\&A, 392, 909
\reference{} Dewi, J., Pols, O.\ 2003, MNRAS, 344, 629
\reference{} Dolan, J.\ 2011, eprint arXiv:1107.1537
\reference{} Dominik, M., Belczynski, K., Fryer, C., Holz, D., Berti, E.,
             Bulik, T., Mandel, I., \& O'Shaughnessy, R.\ 2012, ApJ, submitted 
             (arXiv:1202.4901))
\reference{} Farrell, S., Webb, N., Barret, D., Godet, O., \& Rodrigues, J.\ 2009,
             Nature 460, 73
\reference{} Farr, W., et al.\ 2011, ApJ, accepted (arXiv:1011.1459) 
\reference{} Foglizzo, T.\ 2009, ApJ, 694, 820
\reference{} Fryer, C., \& Kalogera, V.\ 2001, ApJ, 554, 548
\reference{} Fryer, C., \& Warren, M.\ 2004, ApJ, 601, 391
\reference{} Fryer, C., \& Young, P.\ 2007, ApJ, 659, 1438
\reference{} Fryer, C., et al.\ 2012, ApJ, 749, 91
\reference{} Herant, M., et al.\ 1994, ApJ, 435, 339
\reference{} Hobbs, G., et al.\ 2005, MNRAS, 360, 974
\reference{} Hurley, J., Pols, O., \& Tout, C.\ 2000, MNRAS, 315, 543
\reference{} Kreidberg, L., Bailyn, C., Farr, W., \& Kalogera, V.\ 2012, 
             ApJ, submitted (arXiv:1205.1805) 
\reference{} Kroupa, P., \& Weidner, C.\ 2003, ApJ, 598, 1076
\reference{} Liu, Q., van Paradijs, J., van den Heuvel, E.\ 2006,
             A\&A, 455, 1165
\reference{} Lorimer, D., \& McLaughlin, M.\ 2009, Highlights of Astronomy,
             Vol.14 XXVI IAU General Assembly, Corbett, I. ed. (arXiv:1007.3545)
\reference{} Marek, A., Janka, H.-Th. 2009, ApJ, 694, 664 
\reference{} Mirabel, F., \& Rodrigues, I.\ 2003, Science, 300, 1119
\reference{} Murphy, J., Dolence, J., \& Burrows, A.\ 2012, ApJ, submitted
             (arXiv:1205:3491)
\reference{} Nordhaus, J., Burrows, A., Almgren, A., \& Bell, J.\ 2010, ApJ, 720, 694
\reference{} Orosz, J., et al.\ 2011, ApJ, accepted (arXiv:1106.3689)
\reference{} Ozel, F., et al.\ 2010, ApJ 725, 1918
\reference{} Prestwich, A., et al.\ 2007, \apj, 669, L21
\reference{} Rude, G., Orosz, J., McClintock, J., \& Torres, M.\ 2010, 
             Bulletin of the American Astronomical Society, 42, 277
\reference{} Scheck, L., et al.\ 2008, A\&A, 477, 931
\reference{} Slany, P., \& Stuchlik, Z.\ 2008, A\&A 492, 319
\reference{} Sutton, A., Roberts, T., Walton, D., Gladstone, J., \& Scott, 
             A.\ 2012, MNRAS, submitted (arXiv:1203.4100) 
\reference{} Takiwaki, T., Kotake, K., Suwa, Y.\ 2011, ApJ, submitted (arXiv:1108.3989)
\reference{} Tauris, T.M., \& Savonije, G.J.\ 1999, A\&A, 350, 928
\reference{} Timmes, F., Woosley, S., \& Weaver, T.\ 1996, ApJ, 457, 834
\reference{} Ugliano, M., Janka, T., Marek, A., \& Arcones, A.\ 2012, ApJ,
             submitted (arXiv:1205.3657)
\reference{} Vink, J., de Koter, A., \& Lamers, H.\ 2001, \aap, 369, 574
\reference{} Vink, J.S., \& de Koter, A.\ 2002, \aap, 393, 543
\reference{} Webbink, R.\ 1984, ApJ, 277, 355
\reference{} Woosley, S., Heger, A., \& Weaver, T.\ 2002, Rev.Mod.Phys., 74, 1015
\reference{} Ziolkowski, J.\ 2010, Mem. Soc. Astro. Ital., 81, 294 


\end{references}
\end{document}